\documentclass[11pt]{book}
\usepackage{jfaa2e}  
\usepackage{amsmath,amssymb,amsfonts}
\usepackage{latexsym}
\usepackage{graphicx}
\usepackage{epstopdf}

\catcode `\@=11
\@addtoreset{equation}{section}
\def\theequation{\arabic{section}.\arabic{equation}}
\catcode `\@=12
\def\thefigure{\arabic{figure}}

\begin{document}

\author{Y. Wiaux, J. D. McEwen, and P. Vielva}
\communicatedby{Jean-Pierre Antoine}

\chapter{Complex data processing: fast wavelet analysis on the sphere}

\footnotetext{\textit{Math Subject Classifications.}  42C15, 42C40, 65T60, 83F05, 85A40 }
\footnotetext{\textit{Keywords and Phrases.}
                     complex data, wavelet analysis, sphere, cosmology.}

\begin{abstract}
In the general context of complex data processing, this paper reviews a recent
practical approach to the continuous wavelet formalism on the sphere. This formalism notably yields
a correspondence principle which relates wavelets on the plane and on the sphere. Two fast algorithms
are also presented for the analysis of signals on the sphere with steerable wavelets.
\end{abstract}

\section{Introduction\label{sec:Introduction}}

In many fields of science, from computer vision, to biomedical imaging,
geophysics, or astrophysics and cosmology, experiments are set up
releasing more and more complex data to process. A first complexity
of the data lies in their large volume, related to the always increasing
resolution of technological devices. Moreover, data are not necessarily
distributed on the real line (audio signals, ...), or on the plane
(images, ...), but can live on higher-dimensional or nontrivial manifolds
($\mathbb{R}^{n}$, sphere, hyperboloid, ...). Finally, the data may
correspond not only to scalar fields (local intensity), but also to
tensor fields on those manifolds (local diffusion matrix, local polarization,
...).

In this new era of complex data processing, powerful tools always
need to be developed for the precise analysis of the signals under
scrutiny. In this paper, we review recent formal and algorithmic advances
for the continuous wavelet analysis of signals on the sphere. This
scale-space formalism goes well beyond the spectral analysis, as it
enables one to probe the localization, scale, and orientation of the
features of the signals analyzed. An exhaustive review on wavelets
on the sphere and related manifolds is presented in another article
of the present issue \cite{JFAAantoine}.

For the sake of the illustration we choose the example of the cosmic
microwave background (CMB) data from cosmology. The CMB is a polarized
electromagnetic blackbody radiation observed today in all directions
of the sky, which emerged some $380.000$ years after the Big Bang.
This snapshot of the early universe bears a wealth of information
for the study of its structure and evolution, i.e.,  for cosmology.
The present NASA WMAP (\emph{Wilkinson Microwave Anisotropy Probe})
satellite experiment \cite{CMBbennet} releases maps of the celestial
sphere of $3$ megapixels at each detection frequency, while the forthcoming
ESA Planck Surveyor satellite experiment \cite{CMBplanck} will increase
the resolution to $50$ megapixels. The CMB therefore crystallises
the previously quoted potential data complexities. Its temperature
(intensity) and polarization, respectively, define scalar and tensor
fields on the sphere, and the corresponding experimental data already
appear in large volumes. Various applications of the continuous wavelet formalism on the sphere
for the analysis of the CMB are presented in another article
of the present issue \cite{JFAAvielva}.

The structure of the paper goes as follows. We only focus on the formalism
for the continuous wavelet transform
on the sphere introduced in \cite{WSantoine2}, as recently further
developed in a practical approach by \cite{SASwiaux1}. This formalism
is explicitly reviewed in section \ref{sec:Wavelets}.
The wavelet decomposition of a signal on the sphere $S^{2}$ is defined by its projection coefficients on translated,
rotated, and dilated versions of a mother wavelet, i.e.,  a directional
correlation at each analysis scale. These wavelet coefficients therefore
live on the rotation group in three dimensions $SO(3)$. The wavelet
must satisfy an admissibility condition ensuring that the signal may
be explicitly reconstructed from its wavelet coefficients. A correspondence
principle is also recalled stating that wavelets on the sphere may
be built from an inverse stereographic projection of wavelets on the
plane. This principle enables one to transfer onto the sphere some properties
of wavelets on the plane, such as the notion of steerability. We explicitly
describe major examples of axisymmetric, directional, and steerable
wavelets. In section \ref{sec:Directional-correlation}, we give the generic 
definition of directional correlation, and the definition of standard correlation, to which reduce the wavelet
coefficients of a signal with a steerable or axisymmetric wavelet. We discuss their \emph{a priori}
computation cost on any pixelization of $S^{2}$ and of
$SO(3)$, which is prohibitive for high resolution data. We
emphasize the existence of a directional and standard correlation
relation in harmonic space. In section \ref{sec:Fast-algorithms},
we first discuss the band-limitation of signals and filters. We then
review two fast algorithms for the directional correlation of band-limited
signals and filters on iso-latitude
pixelizations on the sphere. The first one is based on a technique
of separation of variables in the Wigner $D$-functions on $SO(3)$
\cite{SASkostelec2,SASwiaux2}. The second one relies on the factorization
of the three-dimensional rotation operators to interpret the result
of the directional correlation as a function on the three-torus $\mathbb{T}^{3}$,
and applies the separation of variables to three-dimensional imaginary
exponentials \cite{SASrisbo,SASwandelt,SASMcewen2}. The \emph{a priori}
$\mathcal{O}(L^{5})$ asymptotic complexity is thereby reduced to
$\mathcal{O}(L^{4})$, where $2L$ roughly stands for the square-root
of the number of pixels on the sphere, i.e.,  for \emph{}band-limited
signals and filters with band-limit $L\in\mathbb{N}$. For steerable
and axisymmetric wavelets, the directional correlation resumes to standard correlations,
and the asymptotic complexity drops to $\mathcal{O}(L^{3})$.
The typical computation time for the directional correlation of megapixels
maps ($L\simeq10^{3}$) correspondingly drops from years to tens of
seconds on a standard computer. This easily allows the analysis of
multiple signals at such high resolutions, and at multiple scales.
These developments finally lead us to our conclusions in section \ref{sec:Conclusion}.

\section{Continuous wavelets on the sphere\label{sec:Wavelets}}

\subsection{Practical approach\label{sub:Practical-approach}}

Among other approaches \cite{WSholschneider,WSfreeden1,WSfreeden2},
a satisfactory formalism for the continuous wavelet transform of signals
on the sphere $S^{2}$ was originally established in a group-theoretical framework
\cite{WSantoine2,WSantoine1,WSantoine3,WSdemanet,WSbogdanova}. 
The aim of the present article is to review a more practical
but completely equivalent approach, recently proposed by \cite{SASwiaux1}.
In that framework, a {}``mother
wavelet'' $\Psi(\omega)$ is defined as a localized square-integrable
function on the unit sphere, on which continuous affine transformations (translations,
rotations, and dilations) may be applied. The wavelet transform of
a square-integrable signal on the sphere is then defined as the directional correlation
of the signal with the dilated versions of the mother wavelet.
At each scale, the corresponding wavelet coefficients are square-integrable
functions on the rotation group in three dimensions $SO(3)$. Finally,
an admissibility condition is imposed on the wavelet which ensures
an exact reconstruction formula of the signal from its wavelet coefficients
\footnote{Notice that the signals and filters considered by the formalism are
scalar functions, i.e.,  invariant under local rotations in the
tangent plane at each point on the sphere. In the general context
of complex data processing, one might want to generalize the wavelet
formalism presented here to the analysis of rank $n$ tensor functions.
However, tensor fields may equivalently be expressed in terms
of scalar fields. In particular, polarization data on the sphere
constitute a rank $2$ tensor field. It can be equivalently described in terms
of its so-called electric and magnetic parts, which actually constitute
two separate functions on the sphere, with a purely scalar behaviour.
The present formalism for the scale-space wavelet decomposition of scalar functions on the
sphere may therefore be applied to the analysis of both scalar fields,
and tensor fields such as polarization data \cite{SASwiaux2,SASwiaux3}.}
.

The real and harmonic structures of the unit sphere $S^{2}$ are concisely summarized as follows.
Any point $\omega$ on the sphere is given in spherical coordinates as $\omega=(\theta,\varphi)$,
in terms of a polar angle, or co-latitude $\theta\in[0,\pi]$, and an azimuthal, or longitudinal angle $ \varphi\in[0,2\pi[ $.
Let $G(\omega)$ be a square-integrable function on the sphere, i.e.,  $G(\omega)$ in $L^{2}(S^{2},d\Omega)$,
with the invariant measure $d\Omega=d\cos\theta d\varphi$. The spherical harmonics form an orthonormal
basis for the decomposition of functions in $L^{2}(S^{2},d\Omega)$.
They are explicitly given in a factorized form in terms of the associated
Legendre polynomials $P_{l}^{m}(\cos\theta)$ and the complex exponentials
$e^{im\varphi}$ as \begin{equation}
Y_{lm}\left(\theta,\varphi\right)=\left[\frac{2l+1}{4\pi}\frac{\left(l-m\right)!}{\left(l+m\right)!}\right]^{1/2}P_{l}^{m}\left(\cos\theta\right)e^{im\varphi},\label{20}\end{equation}
with $l\in\mathbb{N}$, $m\in\mathbb{Z}$, and $|m|\leq l$ \cite{SASabramowitz,SASvarshalovich}. While the index $l$ represents an
overall frequency on the sphere, $\vert m \vert$ represents the frequency associated with the azimuthal variable $ \varphi $.
Any $G(\omega)$ is thus uniquely given as
a linear combination of scalar spherical harmonics
$G\left(\omega\right)=\sum_{l\in\mathbb{N}}\sum_{|m|\leq l}\widehat{G}_{lm}Y_{lm}\left(\omega\right)$ (inverse transform),
for the scalar spherical harmonics coefficients
$\widehat{G}_{lm}=\int_{S^{2}}d\Omega\, Y_{lm}^{*}\left(\omega\right)G\left(\omega\right)$ (direct transform),
with $\vert m\vert\leq l$.

The continuous affine transformations on functions on the sphere are defined as follows.
The operator $R(\omega_{0})$ for the motion, or translation, of amplitude
$\omega_{0}=(\theta_0,\varphi_0)$ of a function reads \begin{equation}
\left[R\left(\omega_{0}\right)G\right]\left(\omega\right)=G\left(R_{\omega_{0}}^{-1}\omega\right),\label{1}\end{equation}
 where $R_{\omega_{0}}(\theta,\varphi)=[R_{\varphi_{0}}^{\hat{z}}R_{\theta_{0}}^{\hat{y}}](\theta,\varphi)$
is defined by the three-dimensional rotation matrices $R_{\theta_{0}}^{\hat{y}}$
and $R_{\varphi_{0}}^{\hat{z}}$, acting on the Cartesian coordinates $(x,y,z)$ in three dimensions
centered on the sphere and associated with $\omega=(\theta,\varphi)$. The rotation operator $R^{\hat{z}}(\chi)$ of a function
around itself, by an angle $\chi\in[0,2\pi[$, is given as \begin{equation}
\left[R^{\hat{z}}\left(\chi\right)G\right]\left(\omega\right)=G\left({R_{\chi}^{\hat{z}}}^{-1}\omega\right),\label{2}\end{equation}
 where $R_{\chi}^{\hat{z}}(\theta,\varphi)=(\theta,\varphi+\chi)$
also follows from the action of the three-dimensional rotation matrix
$R_{\chi}^{\hat{z}}$ on the Cartesian coordinates $(x,y,z)$
associated with $\omega=(\theta,\varphi)$.
The dilation operator $D(a)$ on functions in $L^{2}(S^{2},d\Omega)$,
for a dilation factor $a\in\mathbb{R}_{+}^{*}$, is defined in terms
of the inverse of the corresponding dilation $D_{a}$ on points in
$S^{2}$ as \begin{equation}
\left[D\left(a\right)G\right]\left(\omega\right)=\lambda^{1/2}\left(a,\theta\right)G\left(D_{a}^{-1}\omega\right),\label{3}\end{equation}
with $\lambda^{1/2}(a,\theta)=a^{-1}[1+\tan^{2}(\theta/2)]/[1+a^{-2}\tan^{2}(\theta/2)]$.
The dilated point is given by $D_{a}(\theta,\varphi)=(\theta_{a}(\theta),\varphi)$
with the linear relation $\tan(\theta_{a}(\theta)/2)=a\tan(\theta/2)$.
The dilation operator therefore maps the sphere without its South
pole on itself: $\theta_{a}(\theta):\theta\in[0,\pi[\rightarrow\theta_{a}\in[0,\pi[$.
This dilation operator is uniquely defined by the requirement of the
following natural properties \cite{SASwiaux1}. The dilation $D_{a}$
of points on $S^{2}$ must be a radial (i.e.,  only affecting
the radial variable $\theta$ independently of $\varphi$, and leaving
$\varphi$ invariant) and conformal (i.e.,  preserving the measure
of angles in the tangent plane at each point of $S^{2}$) diffeomorphism
(i.e.,  a continuously differentiable bijection on $S^{2}$). The factor
$\lambda(a,\theta)$ explicitly appears in the conformal transformation of
the metric through the dilation $D_{a}$. The normalization by $\lambda^{1/2}(a,\theta)$
in (\ref{3}) is uniquely determined by the requirement that the dilation $D(a)$ of functions
in $L^{2}(S^{2},d\Omega)$ be a unitary operator (i.e.,  preserving
the scalar product in $L^{2}(S^{2},d\Omega)$, and specifically the
norm of functions).

The analysis of signals goes as follows. The wavelet transform of
a signal $F(\omega)$ in $L^{2}(S^{2},d\Omega)$ on the sphere, with
the wavelet $\Psi(\omega)$, localized analysis function in $L^{2}(S^{2},d\Omega)$,
is defined as the directional correlation between $F(\omega)$ and
the dilated wavelet $\Psi_{a}=D(a)\Psi$, i.e.,  as the scalar
product: \begin{equation}
W_{\Psi}^{F}\left(\rho,a\right)=\langle\Psi_{\rho,a}|F\rangle=\int_{S^{2}}d\Omega\,\Psi_{\rho,a}^{*}\left(\omega\right)F\left(\omega\right),\label{4}\end{equation}
 with $\Psi_{\rho,a}=R(\rho)\Psi_{a}$, and $\rho=(\theta_0,\varphi_0,\chi)$. At each scale, the wavelet
coefficients $W_{\Psi}^{F}(\rho,a)$ are therefore square-integrable
functions on the rotation group in three dimensions $SO(3)$.
They represent the characteristics of the signal for each analysis
scale $a$, direction $\chi$, and position $\omega_{0}$. This defines
the scale-space nature of the wavelet decomposition on the sphere.

The real and harmonic structures of the rotation group
in three dimensions $SO(3)$ are concisely summarized as follows.
Any rotation $\rho$ on $SO(3)$ is given in terms of the three Euler angles $\rho=(\theta,\varphi,\chi)$,
with $\theta\in[0,\pi]$, and $ \varphi,\chi \in[0,2\pi[ $.
Let $H(\rho)$ be a square-integrable function on $SO(3)$, i.e.,  $H(\rho)$ in $L^{2}(SO(3),d\rho)$,
with the invariant measure $d\rho=d\varphi d\cos\theta d\chi$. The Wigner $D$-functions are the
matrix elements of the irreducible unitary representations of weight
$l$ of the group in $L^{2}(SO(3),d\rho)$. By the Peter-Weyl theorem
on compact groups, the matrix elements $D_{mn}^{l*}$ also form an
orthogonal basis in $L^{2}(SO(3),d\rho)$. They are explicitly given
in a factorized form in terms of the real Wigner $d$-functions $d_{mn}^{l}(\theta)$
and the complex exponentials, $e^{-im\varphi}$ and $e^{-in\chi}$
, as \begin{equation}
D_{mn}^{l}\left(\varphi,\theta,\chi\right)=e^{-im\varphi}d_{mn}^{l}\left(\theta\right)e^{-in\chi},\label{23}\end{equation}
with $l\in\mathbb{N}$, $m,n\in\mathbb{Z}$, and $|m|,|n|\leq l$
\cite{SASvarshalovich,CVbrink}.  Again, $l$ represents an
overall frequency on $SO(3)$, and $\vert m \vert$ and $\vert n \vert$ the frequencies associated with the variables
$ \varphi $ and $ \chi $,  respectively. Any $H(\rho)$,
such as the wavelet coefficients at each scale of a signal on $S^2$,
is thus uniquely given as a linear combination of Wigner $D$-functions as
$H\left(\rho\right)=\sum_{l\in\mathbb{N}}(2l+1)/8\pi^{2}\sum_{|m|,|n|\leq l}\widehat{H}_{mn}^{l}D_{mn}^{l*}\left(\rho\right)$ (inverse transform),
with, for $\vert m\vert,\vert n\vert\leq l$, the Wigner $D$-functions coefficients
$\widehat{H}_{mn}^{l}=\int_{SO(3)}d\rho\, D_{mn}^{l}\left(\rho\right)H\left(\rho\right)$ (direct transform).

The synthesis of a signal $F(\omega)$ from its wavelet coefficients
reads as: \begin{equation}
F\left(\omega\right)=\int_{0}^{+\infty}\frac{da}{a^{3}}\int_{SO(3)}d\rho W_{\Psi}^{F}\left(\rho,a\right)\left[R\left(\rho\right)L_{\Psi}\Psi_{a}\right]\left(\omega\right).\label{5}\end{equation}
 In this relation, the operator $L_{\Psi}$ in $L^{2}(S^{2},d\Omega)$
is defined
\footnote{The operator $L_{\Psi}$ in our notations coincides with the inverse of the standard frame operator $A_{\Psi}$ defined in \cite{JFAAantoine}.}
by the following action on the spherical harmonics coefficients
of functions: $\widehat{L_{\Psi}G}_{lm}=\widehat{G}_{lm}/C_{\Psi}^{l}$,
with $|m|\leq l$. This
exact reconstruction formula holds if and only if the spherical harmonics
transform $\widehat{\Psi}_{lm}$ of the wavelet $\Psi(\omega)$ satisfies
the following admissibility condition \cite{SASwiaux1}: \begin{equation}
0<C_{\Psi}^{l}=\frac{8\pi^{2}}{2l+1}\sum_{|m|\leq l}\int_{0}^{+\infty}\frac{da}{a^{3}}\,|\widehat{\left(\Psi_{a}\right)}_{lm}|^{2}<\infty,\label{6}\end{equation}
 for all $l\in\mathbb{N}$. This condition intuitively requires that the whole wavelet family {$\Psi_a(\omega)$},
 for $a\in\mathbb{R}_{+}^{*}$, covers each frequency index $l$ with a finite and non-zero amplitude. As explicitly expressed
 in section \ref{sec:Directional-correlation}, the direct Wigner $D$-functions transform of the wavelet
 coefficients of a signal $F$ with $\Psi$ is given as the pointwise product of the spherical harmonics coefficients
 $\widehat{F}_{lm}$ and $\widehat{(\Psi_a)}_{ln}^{*}$. The admissibility condition consequently requires that the wavelet
 family as a whole preserves the signal information at each frequency $l$.

\subsection{Correspondence principle\label{sub:Correspondence-principle}}

Wavelets on the plane are well-known, and may be easily constructed
as the corresponding admissibility condition reduces to a zero-mean
condition for a function both integrable and square-integrable. On
the contrary, the admissibility condition (\ref{6}) for wavelets
on the sphere is difficult to check in practice. In that context,
a correspondence principle was proved in \cite{SASwiaux1}, stating
that the inverse stereographic projection of a wavelet on the plane
leads to a wavelet on the sphere. 

\begin{figure}[h]
\begin{center}\includegraphics[%
  width=8cm]{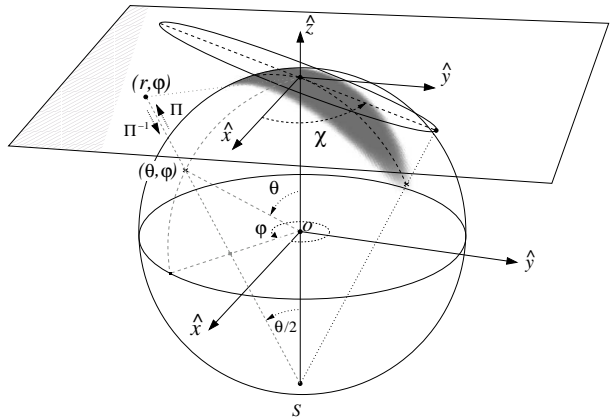}\end{center}

\caption{\label{cap:Stereographic-projection}Stereographic projection $\pi$
and its inverse $\pi^{-1}$, relating points $(\theta,\varphi)$ on
the sphere and $(r,\varphi)$ on its tangent plane at the North pole.
The same relation holds through $\Pi$ and $\Pi^{-1}$ between functions
living on each of the two manifolds, as illustrated by the shadow
on the sphere and the localized region on the plane \cite{SASwiaux1}.}
\end{figure}

The stereographic projection is the unique radial conformal diffeomorphism
mapping the sphere $S^{2}$ onto the plane $\mathbb{R}^{2}$. The
unitary stereographic projection operator between functions $G$ in
$L^{2}(S^{2},d\Omega)$ and $g$ in $L^{2}(\mathbb{R}^{2},d^{2}\vec{x})$,
and its inverse, respectively read \begin{eqnarray}
\left[\Pi G\right]\left(\vec{x}\right) & = & \left(1+\left(\frac{r}{2}\right)^{2}\right)^{-1}G\left(\pi^{-1}\vec{x}\right)\nonumber \\
\left[\Pi^{-1}g\right]\left(\omega\right) & = & \left(1+\tan^{2}\frac{\theta}{2}\right)g\left(\pi\omega\right),\label{7}\end{eqnarray}
 in spherical coordinates on the sphere $\omega=(\theta,\varphi)$,
and polar coordinates on the plane $\vec{x}=(r,\varphi)$. The azimuthal
coordinates on the plane and on the sphere are identified to one another:
$\varphi$. The radial conformal diffeomorphism between points is
given as $\pi(\theta,\varphi)=(r(\theta),\varphi)$ for $r(\theta)=2\tan(\theta/2)$,
and its inverse reads $\pi^{-1}(r,\varphi)=(\theta(r),\varphi)$ for
$\theta(r)=2\arctan(r/2)$. The diffeomorphism $r(\theta)$ and its
inverse $\theta(r)$ explicitly define the stereographic projection
and its inverse. This stereographic projection maps the sphere, without
its South pole, on the entire plane: $r(\theta):\theta\in[0,\pi[\rightarrow[0,\infty[$.
Geometrically, it projects a point $\omega=(\theta,\varphi)$ on the
sphere onto a point $\vec{x}=(r,\varphi)$ on the tangent plane at
the North pole, co-linear with $\omega$ and the South pole (see Fig.
\ref{cap:Stereographic-projection}). The pre-factors in (\ref{7})
are required to ensure the unitarity of the projection operators $\Pi$
and $\Pi^{-1}$.

In this framework, the correspondence principle established states
that, if the function $\psi(r,\varphi)$ in $L^{2}(\mathbb{R}^{2},d^{2}\vec{x})$
satisfies the wavelet admissibility condition on the plane, i.e., 
essentially a zero-mean condition, then the function\begin{equation}
\Psi\left(\theta,\varphi\right)=\left[\Pi^{-1}\psi\right]\left(\theta,\varphi\right),\label{8}\end{equation}
in $L^{2}(S^{2},d\Omega)$, satisfies the wavelet admissibility condition (\ref{6}) on the sphere.
This enables the construction of wavelets on the sphere by
projection of wavelets on the plane. It also transfers
wavelet properties from the plane onto the sphere, such as the
steerability discussed in the next subsection.

\subsection{Axisymmetric, directional, and steerable wavelets\label{sub:Axisymmetric,-directional,-and}}

We present here axisymmetric, directional, and steerable wavelets
on the sphere, built as inverse stereographic projections of wavelets
on the plane.

\begin{figure}[h]
\begin{center}\includegraphics[%
  width=2.5cm]{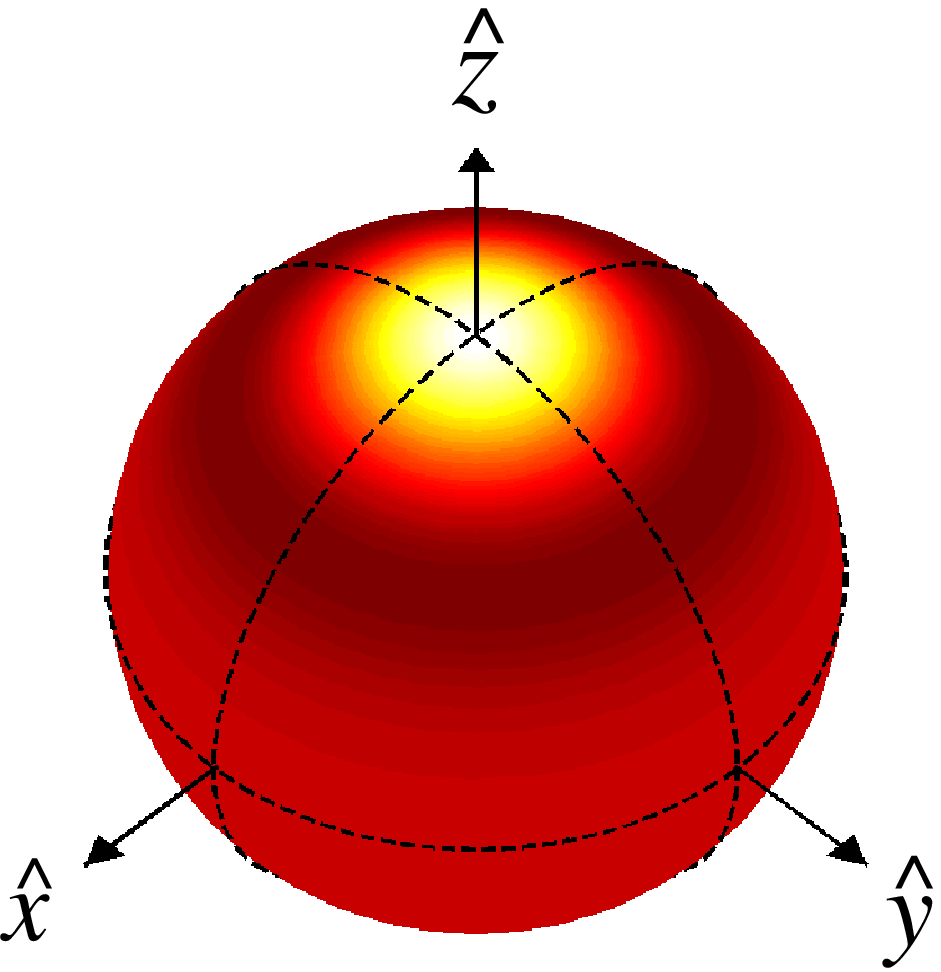}\hspace{2cm}\includegraphics[%
  width=2.5cm]{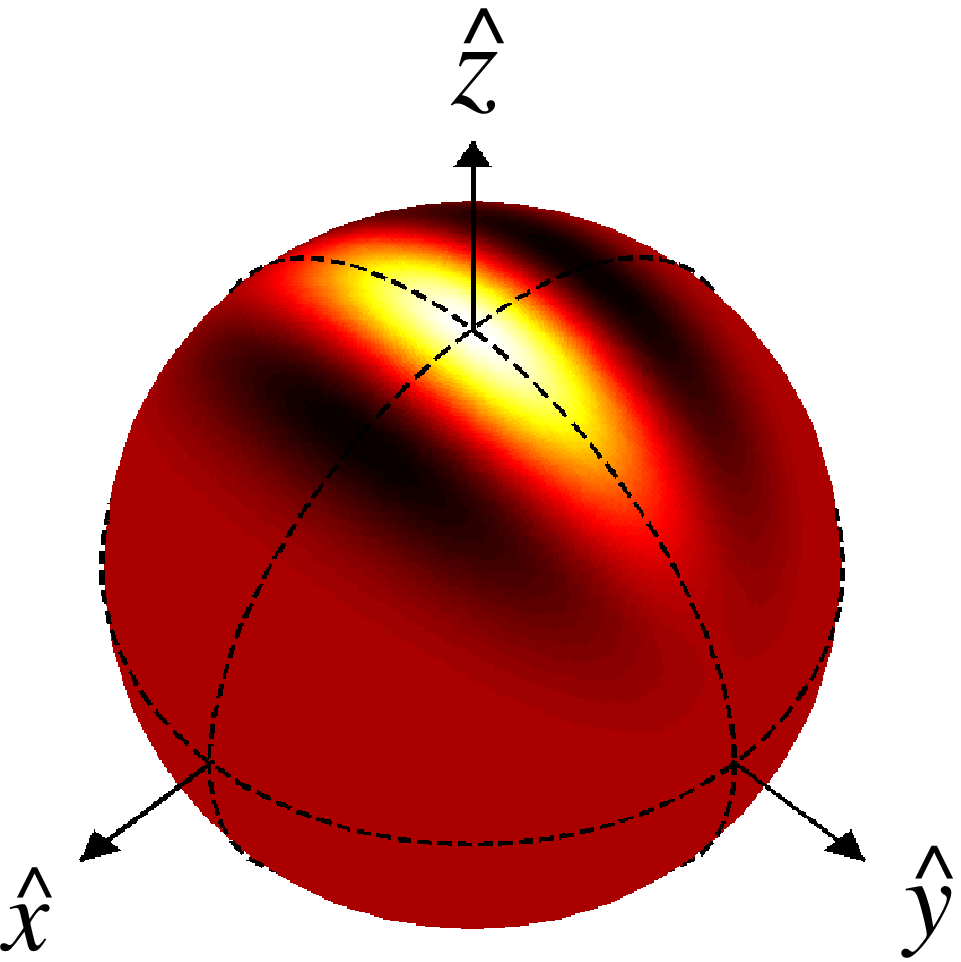}\hspace{2cm}\includegraphics[%
  width=2.5cm]{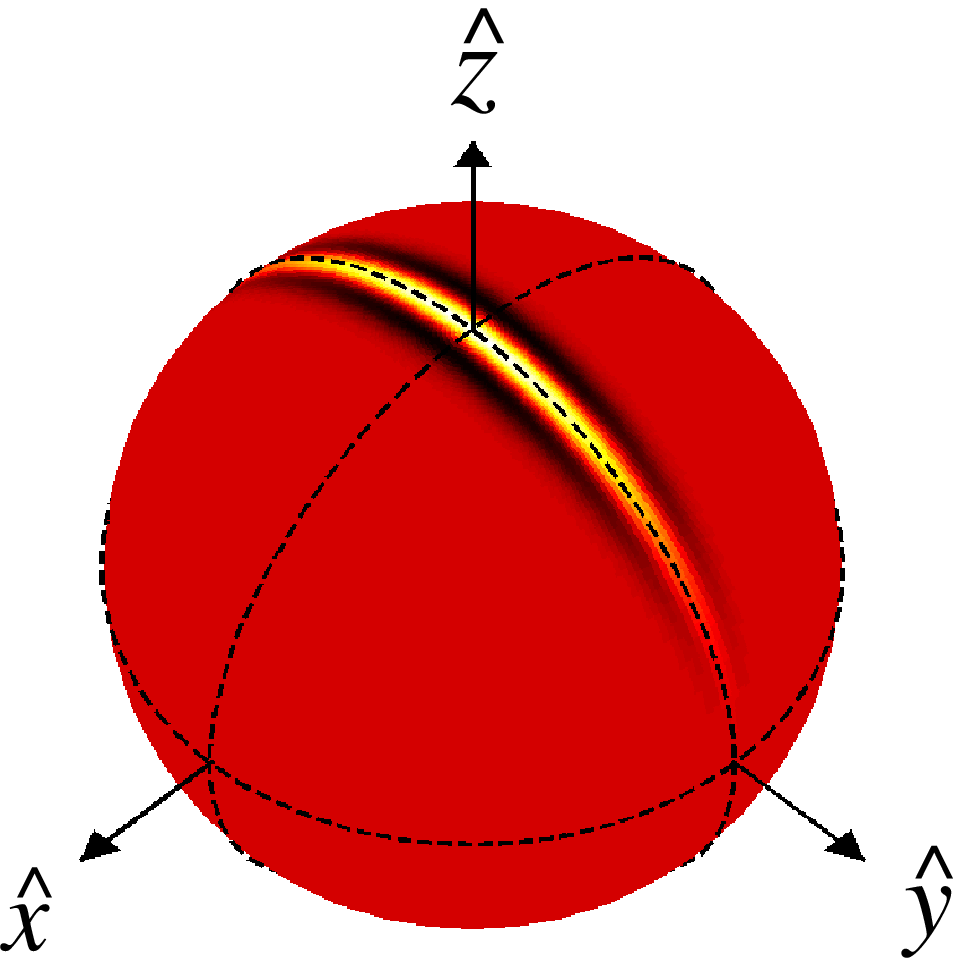}\end{center}

\caption{\label{cap:mex-ell-s2}Mexican hat wavelet on the sphere for a dilation
factor $a=0.4$ and different eccentricities. On the left, the axisymmetric Mexican hat: $r=1$ ($\epsilon=0$)
and $s=2$ (left). At the center and on the right respectively, the elliptical Mexican hat
for $r=0.5$ ($\epsilon\simeq0.96825$) and $s=2$, and $r=0.1$
($\epsilon=0.99995$) and $s=2$. Dark and light regions respectively
identify negative and positive values.}
\end{figure}

An axisymmetric filter is by definition invariant under rotation around
itself. That is, when located at the North pole, an axisymmetric filter
is defined by a function $A(\theta)$ independent of the azimuthal
angle $\varphi$. On the plane, the Mexican hat wavelet is defined
as the normalized (negative) Laplacian of a Gaussian $e^{(x^{2}+y^{2})/2}$.
Its inverse stereographic projection defines the Mexican hat
wavelet on the sphere (see Fig. \ref{cap:mex-ell-s2}).

Any non-axisymmetric filter is said to be directional, and is given
as a general function $\Psi(\theta,\varphi)$ in $L^{2}(S^{2},d\Omega)$.
The elliptical Mexican hat wavelet is a directional modification of
the axisymmetric Mexican hat, obtained by considering different widths
$\sigma_{x}$ and $\sigma_{y}$, respectively in the $\hat{x}$ and
$\hat{y}$ directions on the plane for the original Gaussian: $e^{(x^{2}+y^{2})/2}\rightarrow e^{(x^{2}/\sigma_{x}^{2}+y^{2}/\sigma_{y}^{2})/2}$
\cite{WNGmcewen1}. The wavelet obtained as inverse stereographic
projection of the normalized (negative) Laplacian of this Gaussian
reads (see Fig. \ref{cap:mex-ell-s2}) as:\begin{eqnarray}
\Psi^{(mex)}\left(\omega\right) & = & \sqrt{\frac{2}{\pi}}N\left(\sigma_{x},\sigma_{y}\right)\left(1+\tan^{2}\frac{\theta}{2}\right)\left[1-\frac{4\tan^{2}\theta/2}{\sigma_{x}^{2}+\sigma_{y}^{2}}\right.\nonumber \\
 & & \left.\left(\frac{\sigma_{y}^{2}}{\sigma_{x}^{2}}\cos^{2}\varphi+\frac{\sigma_{x}^{2}}{\sigma_{y}^{2}}\sin^{2}\varphi\right)\right]e^{-2\tan^{2}\frac{\theta}{2}\left(\cos^{2}\varphi/\sigma_{x}^{2}+\sin^{2}\varphi/\sigma_{y}^{2}\right)}.\nonumber \\
 & & \label{11}\end{eqnarray}
The constant $N(\sigma_{x},\sigma_{y})=(\sigma_{x}^{2}+\sigma_{y}^{2})[\sigma_{x}\sigma_{y}(3\sigma_{x}^{4}+3\sigma_{y}^{4}+2\sigma_{x}^{2}\sigma_{y}^{2})/2]^{-1/2}$
stands for the normalization. One can identify the wavelet parameters
through the eccentricity of the ellipse defined by the points where
the wavelet has zero value (zero-crossing), $\epsilon=(1-(\sigma_{x}/\sigma_{y})^{4})^{1/2}$
(for $\sigma_{y}\geq\sigma_{x}$), and the sum $s=\sigma_{x}^{2}+\sigma_{y}^{2}$.
It is alternatively described by the ratio of the semi-major and semi-minor
axes of the Gaussian $r=\sigma_{x}/\sigma_{y}$, and the sum $s=\sigma_{x}^{2}+\sigma_{y}^{2}$.
The axisymmetric Mexican hat is recovered for
$\sigma_{x}=\sigma_{y}=1$, in which case $r=1$ ($\epsilon=0$),
and $s=2$, and the normalization constant is unity, $N(\sigma_{x},\sigma_{y})=1$.

\begin{figure}[h]
\begin{center}\includegraphics[%
  width=2.5cm]{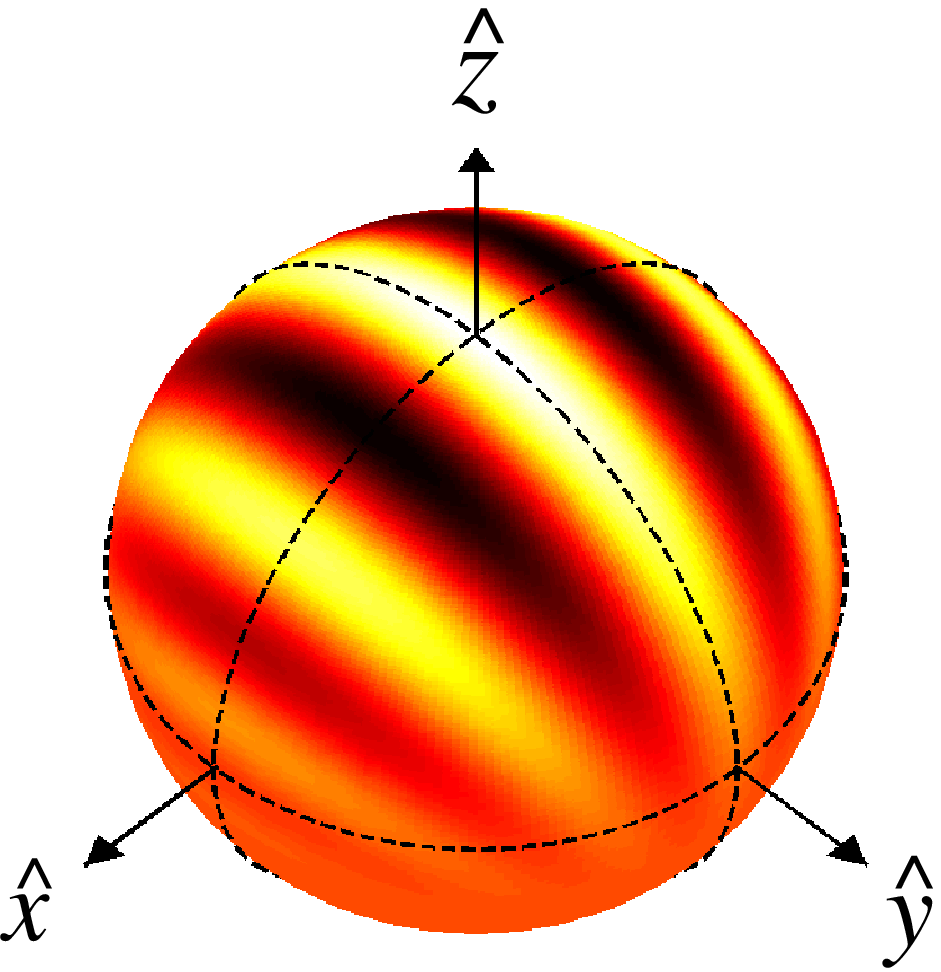}\hspace{2cm}\includegraphics[%
  width=2.5cm]{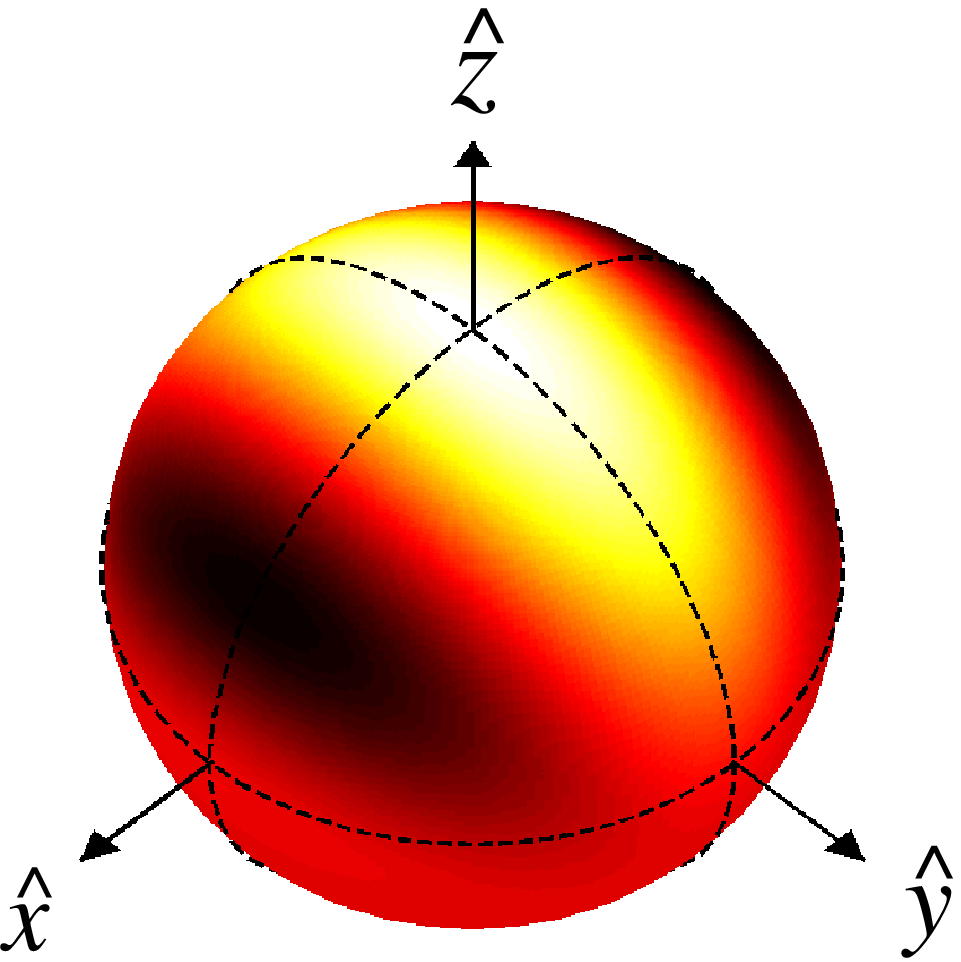}\end{center}

\caption{\label{cap:mor-s2}Real Morlet wavelet on the sphere for a dilation
factor $a=0.4$ and a wave-vector $\vec{k}=(6,0)$ on the left, and for
a dilation factor $a=0.4$ and a wave-vector $\vec{k}=(2,0)$ on the right.
Dark and light regions respectively identify negative and positive
values.}
\end{figure}

On the plane, the real Morlet wavelet is a typical example of a directional
wavelet. Its inverse stereographic projection on the sphere (see Fig.
\ref{cap:mor-s2}) reads as (see also \cite{WSdemanet,WNGmcewen1}
for similar projections):\begin{eqnarray}
\Psi^{(mor)}\left(\omega\right) & = & \sqrt{\frac{2}{\pi}}N\left(k\right)\left(1+\tan^{2}\frac{\theta}{2}\right)\left[\cos\left(\frac{\vec{k}\cdot(\pi^{-1}\vec{x})}{\sqrt{2}}\right)-e^{-\vec{k}^{2}/4}\right] \nonumber \\
& & e^{-2\tan^{2}(\theta/2)},\label{10}\end{eqnarray}
with $\pi^{-1}\vec{x}=(2\tan(\theta/2)\cos\varphi,2\tan(\theta/2)\sin\varphi)$
in Cartesian coordinates. The arbitrary wave-vector $\vec{k}=(k_{x},k_{y})$
controls the direction and the frequency of oscillation of the wavelet
($\vec{k}^{2}=k_{x}^{2}+k_{y}^{2}$). The constant $N(k)=(1+3e^{-\vec{k}^{2}/2}-4e^{-3\vec{k}^{2}/8})^{-1/2}$
stands for the normalization. Notice that for
$\vec{\vert k}\vert=2$, the real Morlet wavelet closely approximates
at large scales to the second Gaussian derivative described
in the following.

The notion of filter steerability was first introduced on the plane
\cite{STfreeman,STsimoncelli}, and more recently defined on the sphere
\cite{SASwiaux1}. Just as on the plane, a directional filter $\Psi$
in $L^{2}(S^{2},d\Omega)$ on the sphere is steerable if any rotation
by $\chi\in[0,2\pi[$ of the filter around itself $R^{\hat{z}}(\chi)\Psi$
may be expressed as a linear combination of a finite number of basis
filters $\Psi_{m}$:\begin{equation}
\left[R^{\hat{z}}\left(\chi\right)\Psi\right]\left(\omega\right)=\sum_{m=1}^{M}k_{m}\left(\chi\right)\Psi_{m}\left(\omega\right).\label{12}\end{equation}
 The weights $k_{m}(\chi)$, with $1\leq m\leq M$, and $M\in\mathbb{N}$,
are called interpolation functions. In particular cases, the basis
filters may be specific rotations by angles $\chi_{m}$ of the original
filter: $\Psi_{m}=R^{\hat{z}}(\chi_{m})\Psi$. Steerable filters have
a nonzero angular width in the azimuthal angle $\varphi$ which makes
them sensitive to a whole range of directions and enables them to
satisfy the relation (\ref{12}). In the spherical harmonics space,
this nonzero angular width corresponds to an azimuthal angular band
limit $N\in\mathbb{N}$ in the frequency index $n$ associated with the azimuthal
variable $\varphi$: \begin{equation}
\widehat{\Psi}_{ln}=0\quad\textnormal{for}\quad\vert n\vert\geq N.\label{13}\end{equation}
Typically, the number $M$ of interpolating functions is of the same
order as the azimuthal band limit $N$.

The derivatives of order $N_{d}$ in direction $\hat{x}$ of radial functions
on the plane are steerable wavelets. The transfer of the steerability
property (\ref{12}) from the plane to the sphere is obvious since
the inverse stereographic projection is a radial operation, while
the steerability only affects the azimuthal variable. The inverse
stereographic projection of Gaussian derivatives therefore give steerable
wavelets on the sphere. They may be rotated in terms of $M=N_{d}+1$
basis filters, and are band-limited in $\varphi$ at $N=N_{d}+1$.
We give the explicit examples of the normalized first and second Gaussian
derivatives. A first derivative has a band limit $N=2$, and only
contains the frequencies $n=\{\pm1\}$. It may be rotated in terms of
two specific rotations at $\chi=0$ and $\chi=\pi/2$, corresponding
to the inverse projection of the first derivatives in directions $\hat{x}$
and $\hat{y}$, $\Psi^{\partial_{\hat{x}}}$ and $\Psi^{\partial_{\hat{y}}}$
respectively:\begin{equation}
\left[R^{\hat{z}}\left(\chi\right)\Psi^{\partial_{\hat{x}}}\right]\left(\omega\right)=\Psi^{\partial_{\hat{x}}}\left(\omega\right)\cos\chi+\Psi^{\partial_{\hat{y}}}\left(\omega\right)\sin\chi.\label{14}\end{equation}

\begin{figure}[h]
\begin{center}\includegraphics[%
  width=2.5cm]{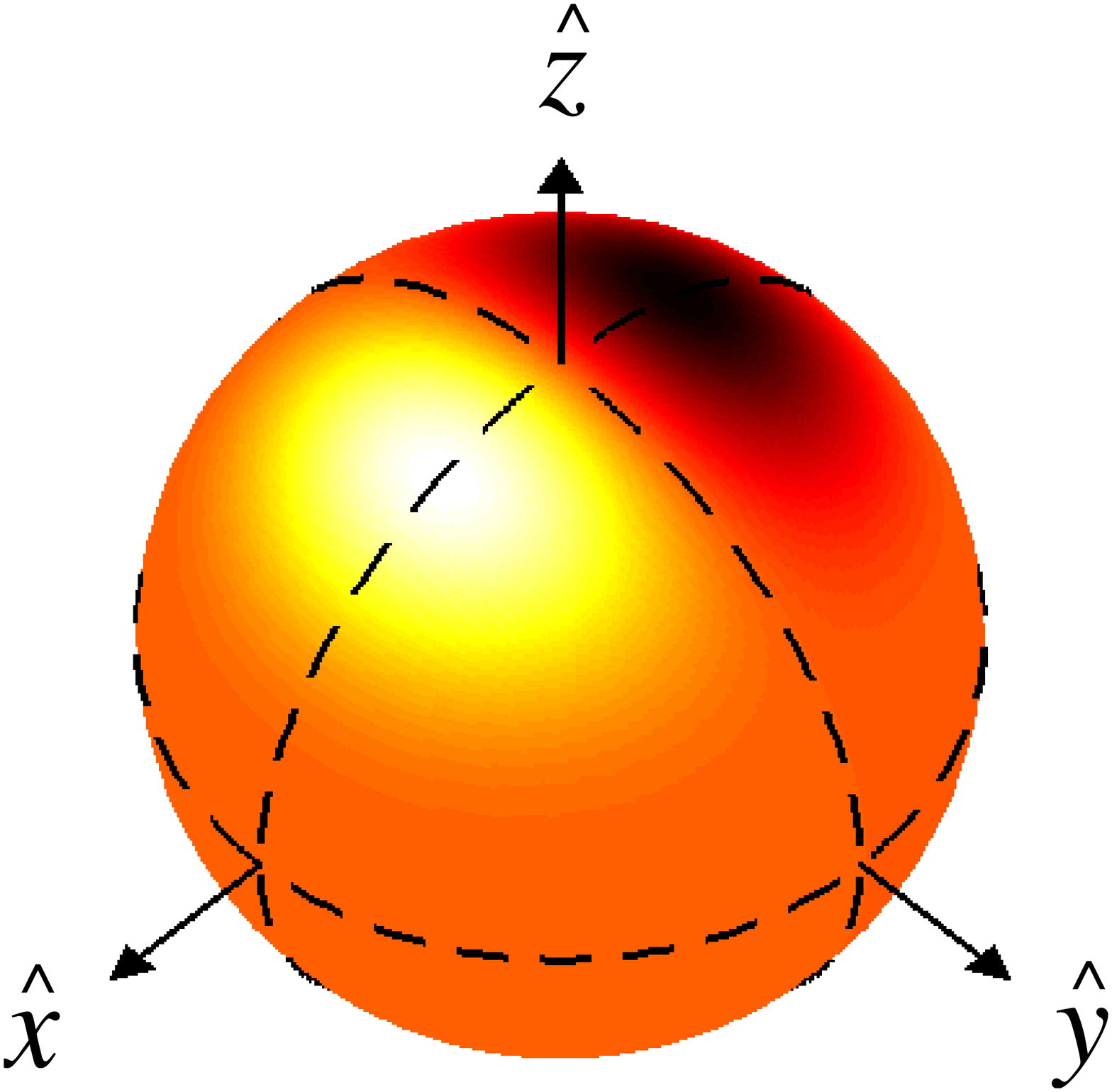}\hspace{1cm}\includegraphics[%
  width=2.5cm]{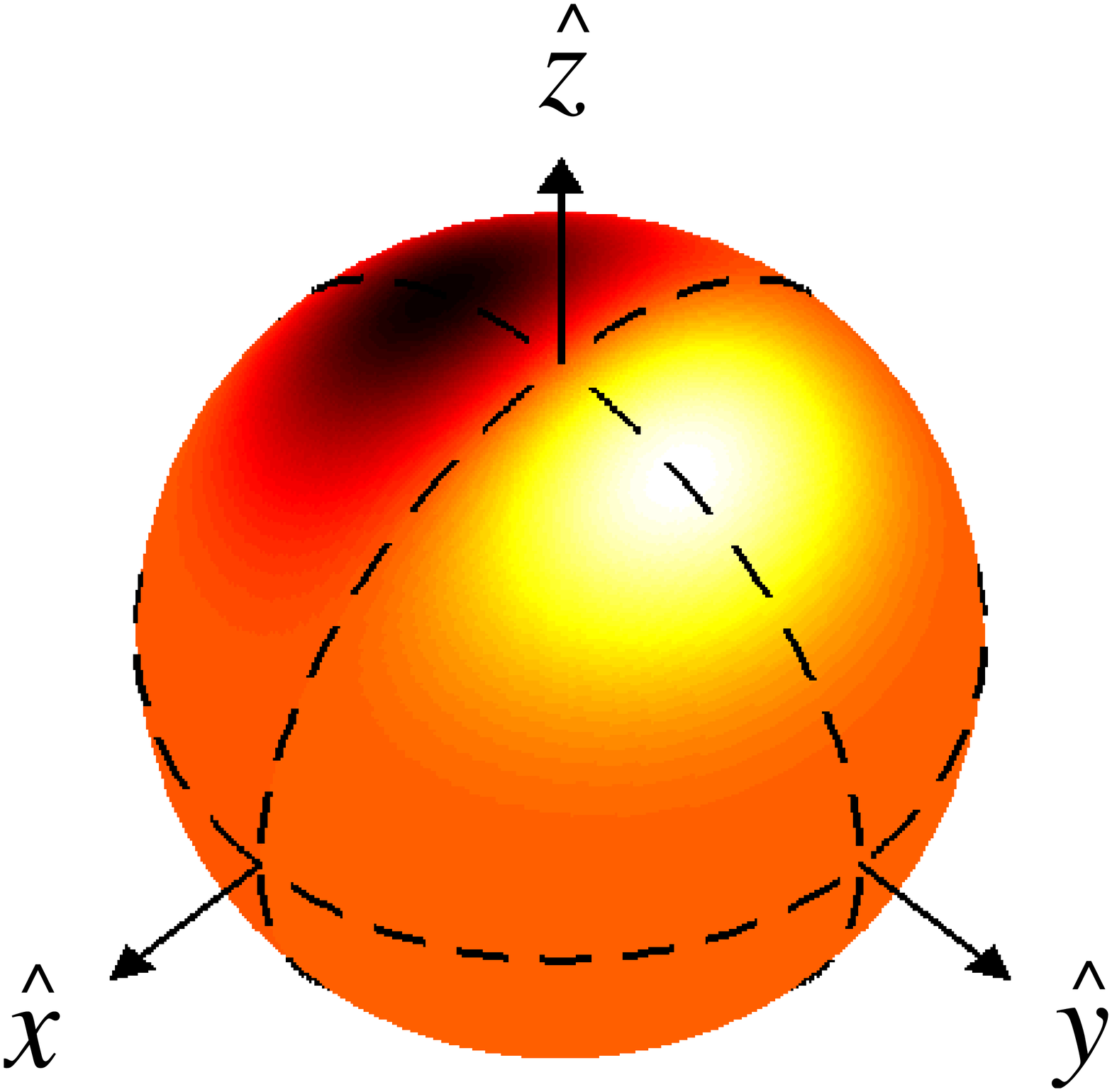}\hspace{1cm},\hspace{1cm}\includegraphics[%
  width=2.5cm]{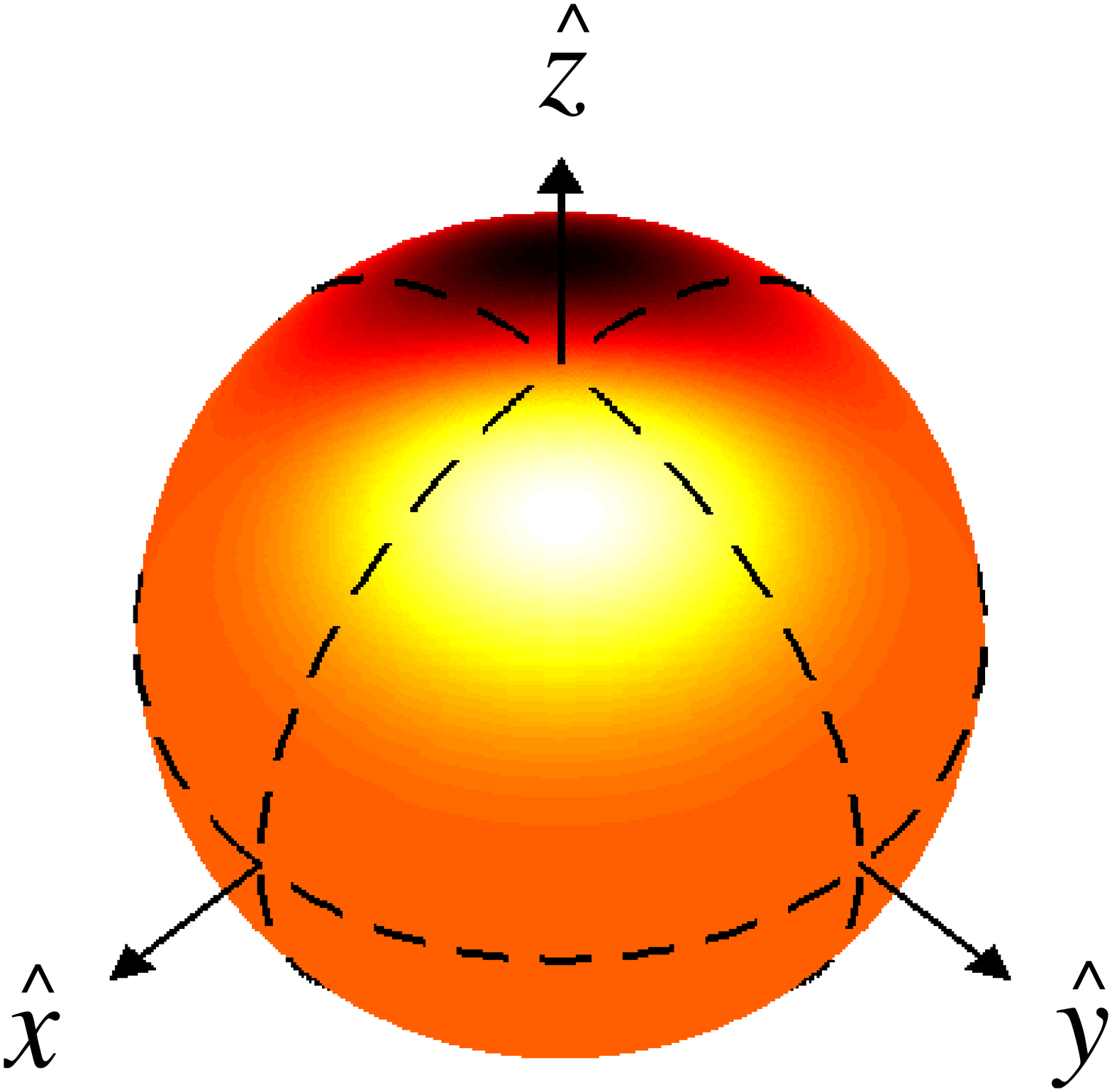}\end{center}

\caption{\label{cap:First-gaussian-steerability-s2}First Gaussian derivative
wavelet on the sphere for a dilation factor $a=0.4$: from left to right, $\Psi^{\partial_{\hat{x}}(gau)}$,
$\Psi^{\partial_{\hat{y}}(gau)}$, and rotation
by $\chi=\pi/4$ of $\Psi^{\partial_{\hat{x}}(gau)}$. Dark and
light regions respectively identify negative and positive values \cite{SASwiaux1}.}
\end{figure}

The normalized first derivatives of a Gaussian (see Fig. \ref{cap:First-gaussian-steerability-s2})
in directions $\hat{x}$ and $\hat{y}$ read:\begin{eqnarray}
\Psi^{\partial_{\hat{x}}(gau)}\left(\theta,\varphi\right) & = & \sqrt{\frac{8}{\pi}}\left(1+\tan^{2}\frac{\theta}{2}\right)\tan\frac{\theta}{2}\cos\varphi e^{-2\tan^{2}(\theta/2)}\nonumber \\
\Psi^{\partial_{\hat{y}}(gau)}\left(\theta,\varphi\right) & = & \sqrt{\frac{8}{\pi}}\left(1+\tan^{2}\frac{\theta}{2}\right)\tan\frac{\theta}{2}\sin\varphi e^{-2\tan^{2}(\theta/2)}.\label{15}\end{eqnarray}

A second derivative has a band limit $N=3$, and contains the frequencies
$n=\{0,\pm2\}$. It may be rotated in terms of three basis filters.
It reads indeed in terms of the inverse projection of the second derivatives
in directions $\hat{x}$ and $\hat{y}$, $\Psi^{\partial_{\hat{x}}^{2}}$
and $\Psi^{\partial_{\hat{y}}^{2}}$ respectively, and the cross derivative
$\Psi^{\partial_{\hat{x}}\partial_{\hat{y}}}$ as:\begin{equation}
\left[R^{\hat{z}}\left(\chi\right)\Psi^{\partial_{\hat{x}}^{2}}\right]\left(\omega\right)=\Psi^{\partial_{\hat{x}}^{2}}\left(\omega\right)\cos^{2}\chi+\Psi^{\partial_{\hat{y}}^{2}}\left(\omega\right)\sin^{2}\chi+\Psi^{\partial_{\hat{x}}\partial_{\hat{y}}}\left(\omega\right)\sin2\chi.\label{16}\end{equation}

\begin{figure}[h]
\begin{center}\includegraphics[%
  width=2.5cm]{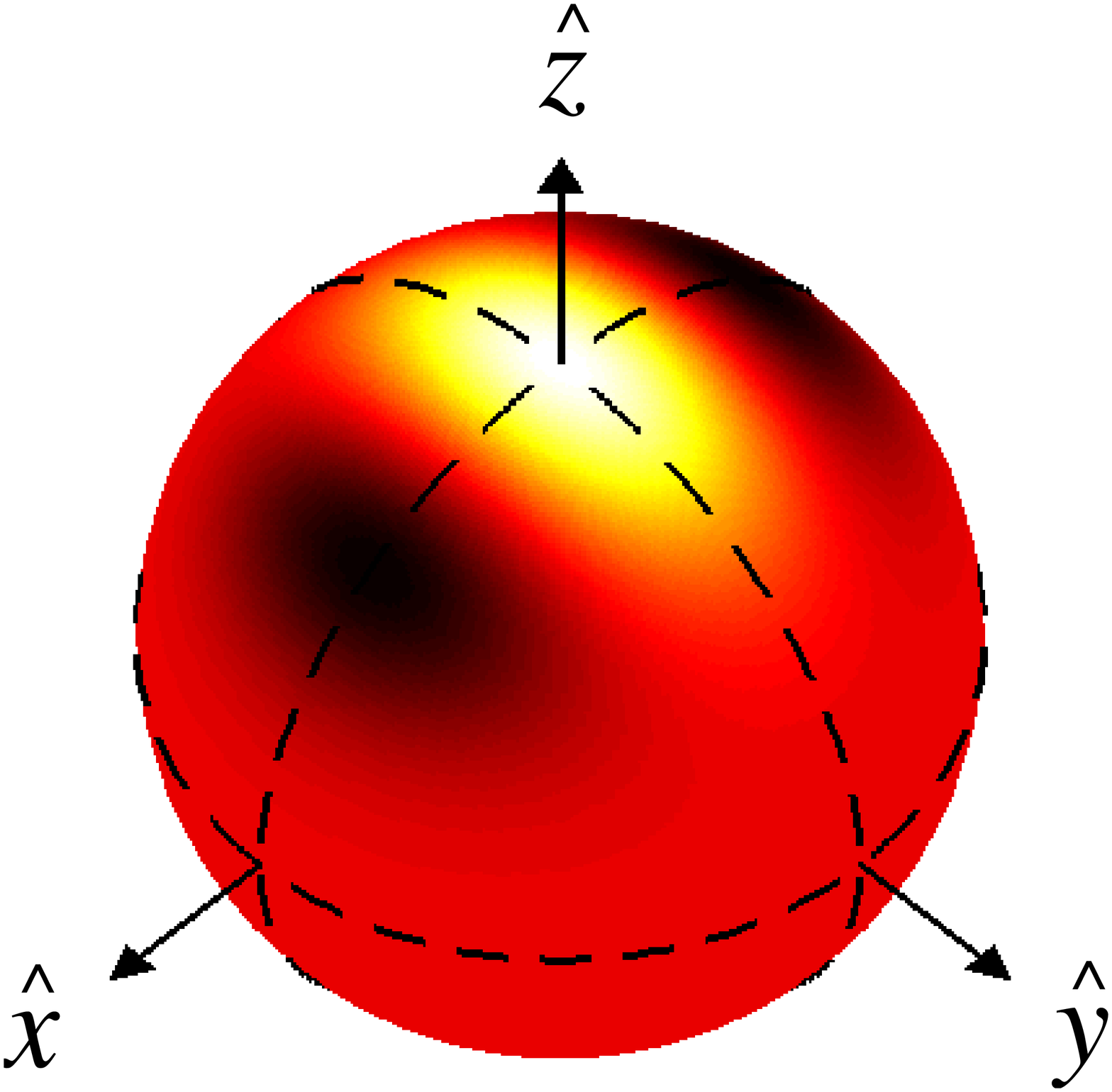}\hspace{0.5cm}\includegraphics[%
  width=2.5cm]{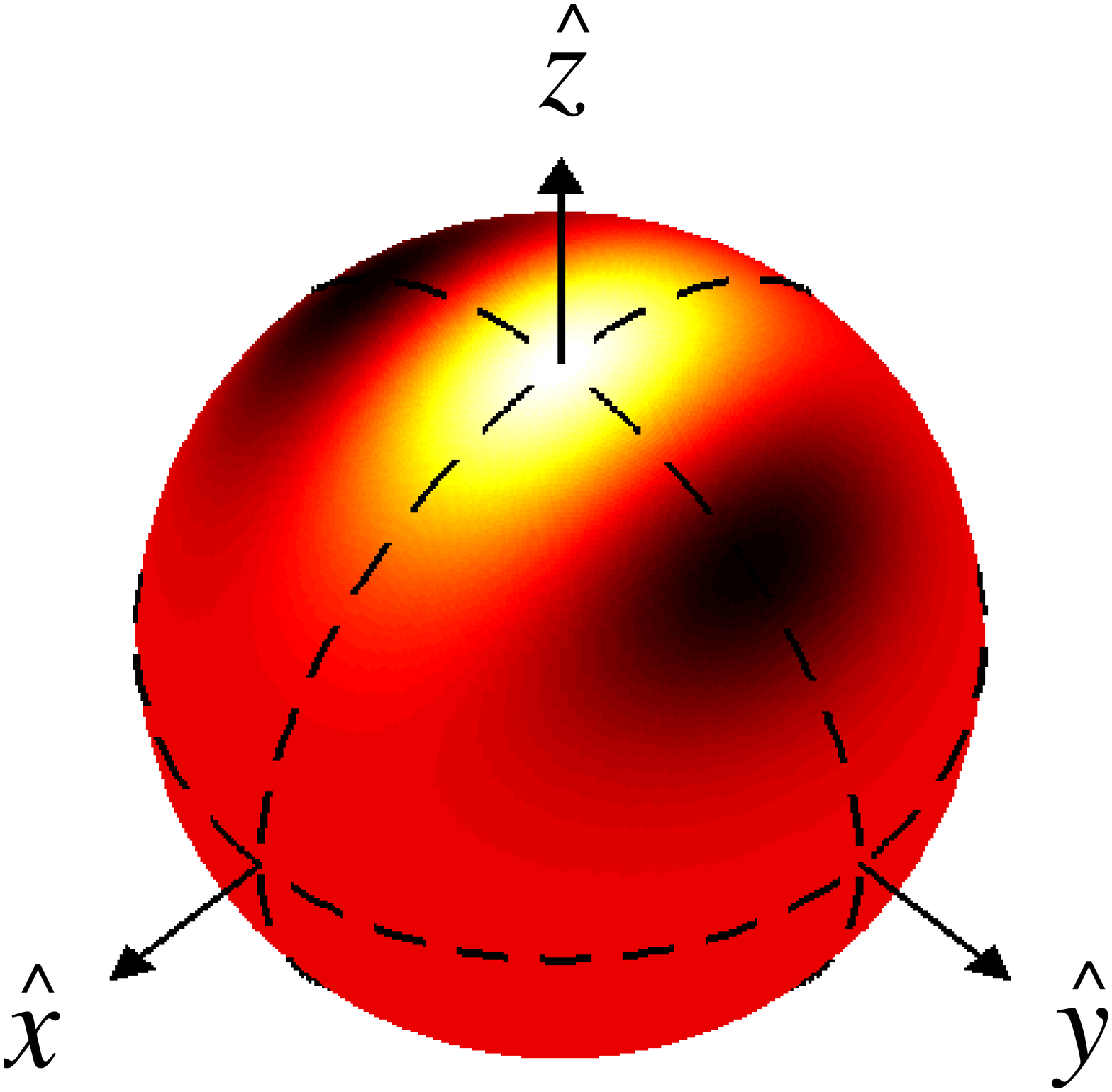}\hspace{0.5cm}\includegraphics[%
  width=2.5cm]{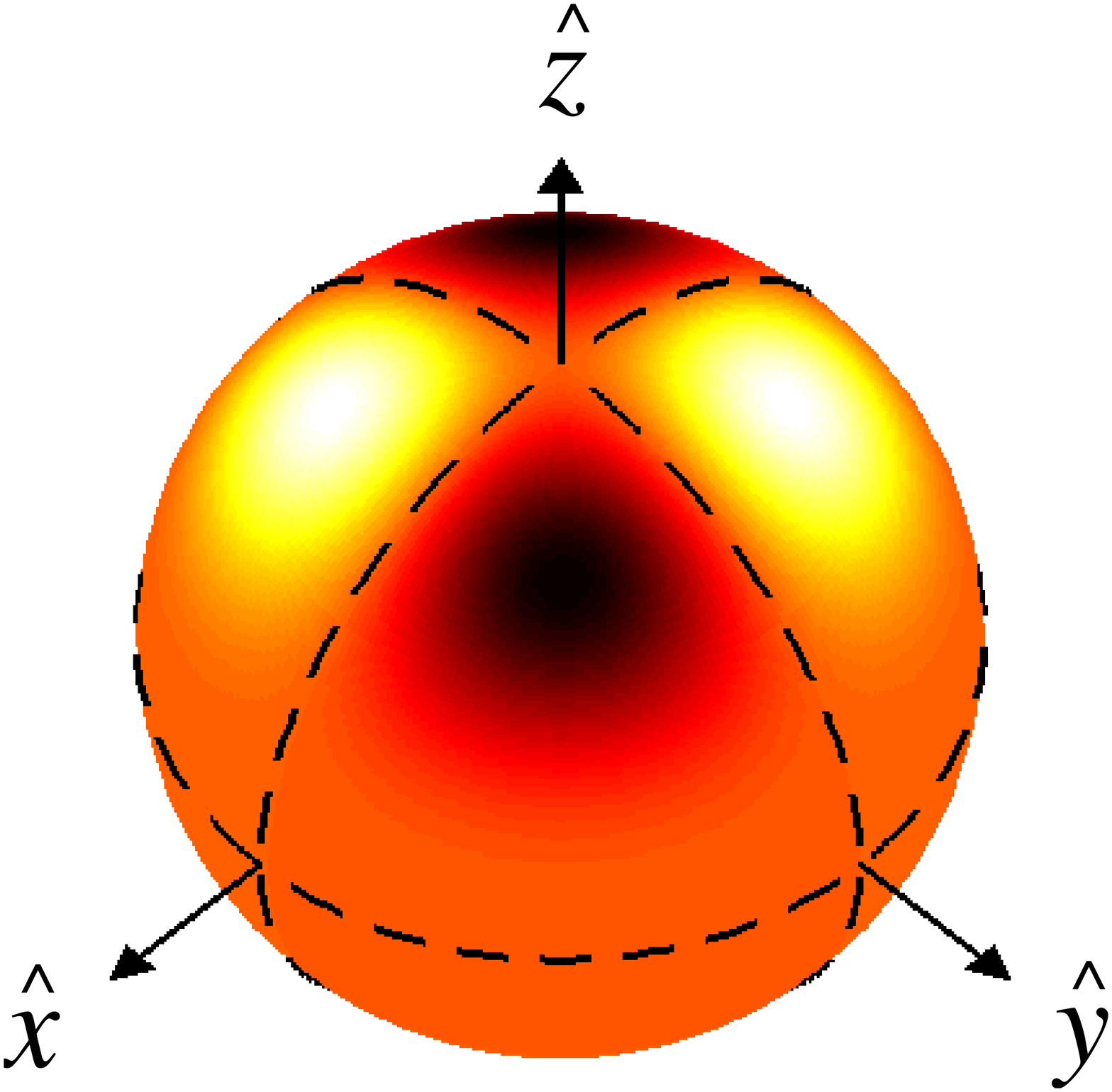}\hspace{0.5cm},\hspace{0.5cm}\includegraphics[%
  width=2.5cm]{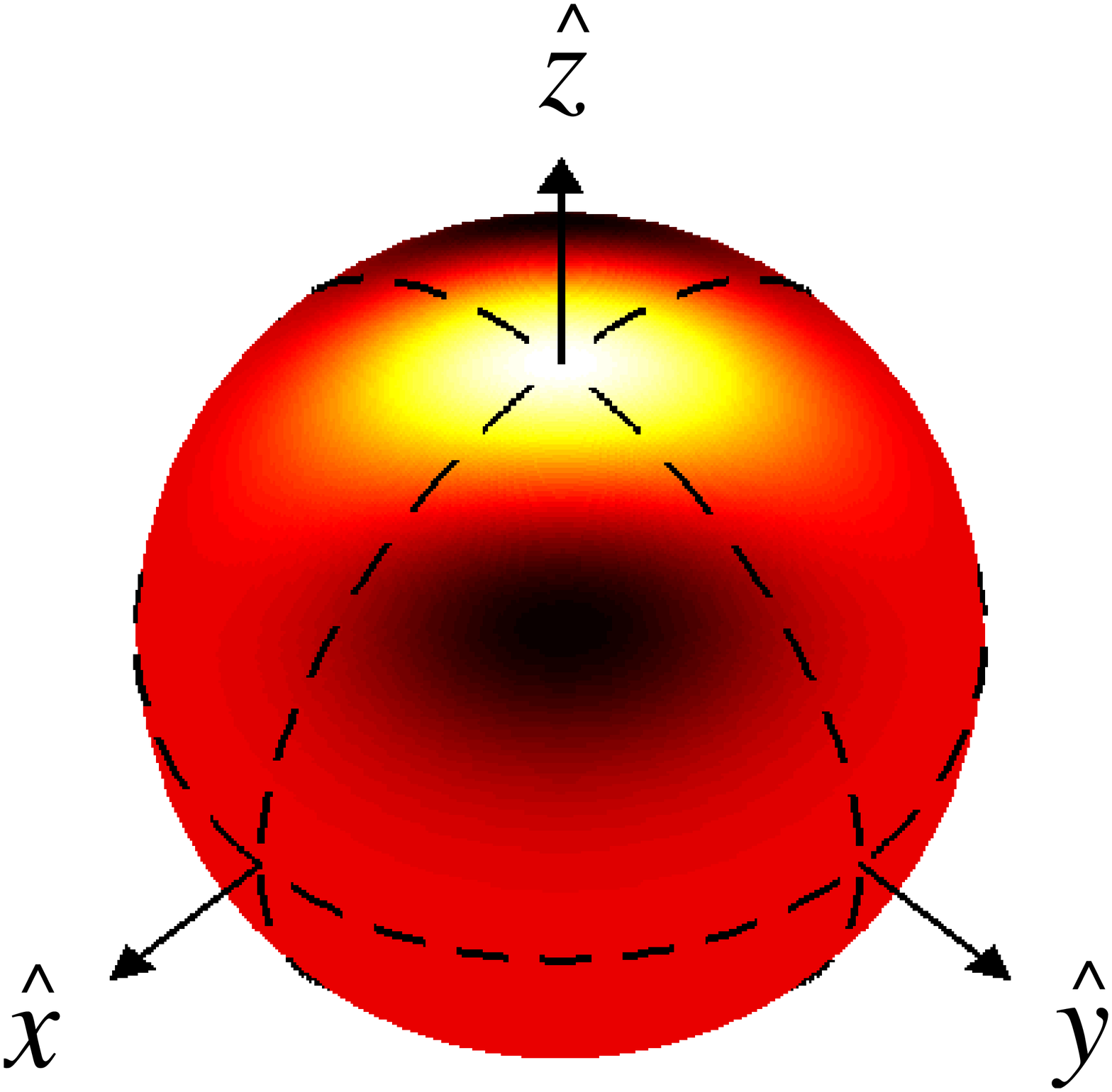}\end{center}

\caption{\label{cap:Second-gaussian-steerability-s2}Second Gaussian derivative
wavelet on the sphere for a dilation factor $a=0.4$: from left to right,
$\Psi^{\partial_{\hat{x}}^{2}(gau)}$,
$\Psi^{\partial_{\hat{y}}^{2}(gau)}$,
$\Psi^{\partial_{\hat{x}}\partial_{\hat{y}}(gau)}$, and
rotation by $\chi=\pi/4$ of $\Psi^{\partial_{\hat{x}}^{2}(gau)}$.
Dark and light regions respectively identify negative and positive
values \cite{SASwiaux1}.}
\end{figure}

The correctly normalized second derivatives of a Gaussian (see Fig.
\ref{cap:Second-gaussian-steerability-s2}) in directions $\hat{x}$
and $\hat{y}$ read:\begin{eqnarray}
\Psi^{\partial_{\hat{x}}^{2}(gau)}\left(\theta,\varphi\right) & = & \sqrt{\frac{4}{3\pi}}\left(1+\tan^{2}\frac{\theta}{2}\right)\left(1-4\tan^{2}\frac{\theta}{2}\cos^{2}\varphi\right)e^{-2\tan^{2}(\theta/2)}\nonumber \\
\Psi^{\partial_{\hat{y}}^{2}(gau)}\left(\theta,\varphi\right) & = & \sqrt{\frac{4}{3\pi}}\left(1+\tan^{2}\frac{\theta}{2}\right)\left(1-4\tan^{2}\frac{\theta}{2}\sin^{2}\varphi\right)e^{-2\tan^{2}(\theta/2)}\nonumber \\
\Psi^{\partial_{\hat{x}}\partial_{\hat{y}}(gau)}\left(\theta,\varphi\right) & = & -\frac{4}{\sqrt{3\pi}}\left(1+\tan^{2}\frac{\theta}{2}\right)\left(\tan^{2}\frac{\theta}{2}\sin2\varphi\right)e^{-2\tan^{2}(\theta/2)}.\nonumber \\
\label{17}\end{eqnarray}

\section{Directional correlation\label{sec:Directional-correlation}}

\subsection{Directional and standard correlations\label{sub:Directional-and-standard}}

The directional correlation $\langle R\Psi\vert F\rangle$ of a function
$F$ with a filter $\Psi$ is generically defined as the scalar product of the
signal with all $SO(3)$ rotations of the filter \cite{SASwiaux2}.
It therefore lives on $SO(3)$, and explicitly reads in $L^{2}(SO(3),d\rho)$
as:

\begin{equation}
\langle R\left(\rho\right)\Psi\vert F\rangle=\int_{S^{2}}d\Omega\,\Psi^{*}\left(R_{\rho}^{-1}\omega\right)F\left(\omega\right).\label{18}\end{equation}
As discussed in subsection \ref{sub:Practical-approach}, if $\Psi$
is the specific dilation of a wavelet on the sphere, the directional
correlation coincides with the wavelet coefficients of the signal,
at the corresponding scale (see relation (\ref{4})).

The standard correlation $\langle R_{0}\Psi\vert F\rangle$ of $F$
with $\Psi$, is generically defined by the scalar product between the function
$F$ and the filter $\Psi$ translated at any point $\omega_{0}=(\theta_{0},\varphi_{0})$
on the sphere, but for a fixed direction, i.e.,  a fixed value
$\chi=0$. The result of the standard correlation explicitly gives
a square-integrable function in $L^{2}(S^{2},d\Omega)$ on the sphere:\begin{equation}
\langle R\left(\omega_{0}\right)\Psi\vert F\rangle=\int_{S^{2}}d\Omega\,\Psi^{*}\left(R_{\omega_{0}}^{-1}\omega\right)F\left(\omega\right).\label{19}\end{equation}
 The notation $R_{0}$ simply denotes a three-dimensional rotation
with $\chi=0$. It distinguishes the standard correlation $\langle R_{0}\Psi\vert F\rangle$
from the directional correlation $\langle R\Psi\vert F\rangle$ when
the arguments are not specified.

Let us remark that, from relation (\ref{12}), it explicitly appears
that the directional correlation with a steerable filter $\Psi$ reduces
to a $M$-terms linear combination of standard correlations with the
corresponding basis filters $\Psi_{m}$. In the particular case of
an axisymmetric filter, there is no dependence at all of the correlation
in the filter rotation $\chi$. The directional correlation with an
axisymmetric filter is therefore strictly equivalent to the standard
correlation.

\subsection{Pixelization and \emph{a priori} computation cost\label{sub:A-priori-computation}}

The directional and standard correlations are defined for square-integrable
functions on a continuous variable $\omega=(\theta,\varphi)$ on the
sphere. The translations and rotations of the filter also form a continuous
variable $\rho=(\varphi_{0},\theta_{0},\chi)\in SO(3)$. Practical
implementations are obviously based on a choice of discretization
for each of these variables, i.e.,  a pixelization of $S^{2}$
and $SO(3)$. Let $N_{p}\simeq(2L)^{2}$ represent the number of sampling
points $\omega$ in a given pixelization of $S^{2}$. The quantity
$2L$ represents the mean number of sampling points in the position
variables $\theta$ and $\varphi$, or $\theta_{0}$ and $\varphi_{0}$.
A simple extrapolation of the Nyquist-Shannon theorem on the line
intuitively associates $L\in\mathbb{N}$ with the band limit, or maximum
frequency, accessible on that pixelization in the {}``Fourier''
indices conjugate to $\theta$ and $\varphi$ for the signals and
filters considered. For a sampling on $2L$ points in the direction
$\chi$ the same band limit $L$ is associated with the conjugate
Fourier index. Notice that in the wavelet formalism, the dilation
parameter $a\in\mathbb{R}_{+}^{*}$ must also be discretized for practical
purposes.

Considering a simple quadrature, i.e.,  a discretization of the
directional correlation integral, each scalar product on the sphere
has an asymptotic complexity $\mathcal{O}(L^{2})$. The overall asymptotic
complexity for the directional correlation (\ref{18}), taking into
account all discrete $\rho=(\varphi_{0},\theta_{0},\chi)$ on $SO(3)$,
is therefore of order $\mathcal{O}(L^{5})$. We consider fine samplings
of several megapixels on the sphere. To fix ideas, let us notice that
the present NASA WMAP precision experiment on the CMB provides maps
of the celestial sphere of around $3$ megapixels. For a sampling
associated with a band limit around $L\simeq10^{3}$, the typical
computation times for $(2L)^{2}$ multiplications and $(2L)^{2}$
additions are of order of $10^{-2}$ seconds on a standard processor.
We take this value as a fair estimation of the computation time required
for a scalar product. Consequently, a unique $\mathcal{O}(L^{5})$
directional correlation would take several years at that band limit
on a single standard computer. Moreover, depending on the application,
the directional correlation of multiple signals might be required.
Typically, thousands of simulated signals are to be considered for
a Monte Carlo statistical analysis. And for a wavelet analysis, multiple
scales are to be considered for the filter. In conclusion, the directional
correlation analysis of functions on the sphere is absolutely unaffordable
for fine samplings with a band limit around $L\simeq10^{3}$ in $\theta$,
$\varphi$, and $\chi$. This conclusion remains when the use of multiple
computers is envisaged. It is even strongly reinforced in the perspective
of an analysis from finer pixelizations on the sphere. In particular,
the forthcoming ESA Planck CMB experiment will provide $50$ megapixels
maps\emph{, i.e., } $L\simeq4\times10^{3}$.

The overall asymptotic complexity for the standard correlation, taking
into account all discrete $\omega_{0}=(\varphi_{0},\theta_{0})$ on
$S^{2}$, is of order $\mathcal{O}(L^{4})$. On a single standard
computer, the corresponding computation time through simple quadrature,
at a band limit $L\simeq10^{3}$, would be of the order of days. Such
a calculation still remains hardly affordable, particularly when multiple
signals and multiple scales are considered.

\subsection{Directional and standard correlations in harmonic space\label{sub:Dir-corr-harmonicspace}}

The Wigner $D$-functions coefficients $\widehat{\langle R\Psi\vert F\rangle}_{mn}^{l}$
of the directional correlation $\langle R(\rho)\Psi\vert F\rangle$
living on $SO(3)$ are given as the pointwise product of the spherical
harmonics coefficients $\widehat{F}_{lm}$ and $\widehat{\Psi}_{ln}^{*}$.
The following correlation relation holds:\begin{equation}
\langle R\left(\rho\right)\Psi\vert F\rangle=\sum_{l\in\mathbb{N}}\frac{2l+1}{8\pi^{2}}\sum_{|m|,|n|\leq l}\widehat{\langle R\Psi\vert F\rangle}_{mn}^{l}D_{mn}^{l*}\left(\rho\right),\label{26}\end{equation}
with\begin{equation}
\widehat{\langle R\Psi\vert F\rangle}_{mn}^{l}=\frac{8\pi^{2}}{2l+1}\widehat{\Psi}_{ln}^{*}\widehat{F}_{lm}.\label{27}\end{equation}
Indeed, the orthonormality of scalar spherical harmonics implies the
Plancherel relation $\langle R\Psi\vert F\rangle=\sum_{l\in\mathbb{N}}\sum_{|m|\leq l}\widehat{R\Psi}_{lm}^{*}\widehat{F}_{lm}$.
The action of the operator $R(\rho)$ on a function $G(\omega)$ in
$L^{2}(S^{2},d\Omega)$ on the sphere reads in terms of its spherical
harmonics coefficients as: $\widehat{[R(\rho)G]}_{lm}=\sum_{|n|\leq l}D_{mn}^{l}(\rho)\widehat{G}_{ln}.$
Inserting this last relation for $\Psi$ in the former Plancherel
relation finally gives the result.

The standard correlation $\langle R(\omega_{0})\Psi\vert F\rangle$
lives on $S^{2}$ and could be decomposed in its spherical harmonics
coefficients. However, for non-axisymmetric filters, these coefficients
do not appear as a simple pointwise product similar to (\ref{27}).
The easiest way to express the standard correlation in harmonic space
is therefore to simply consider the relations (\ref{26}) and (\ref{27})
with $\chi=0$.

\section{Fast algorithms\label{sec:Fast-algorithms}}

\subsection{Band-limitation\label{sub:Band-limitation}}

The wavelet formalism defined in section \ref{sec:Wavelets}
holds for any signal and any wavelet satisfying the admissibility condition
(\ref{6}), irrespectively of any band-limitation consideration. However, the band-limitation
represents a necessary condition for obtaining precise numerical implementations.
We therefore consider band-limited functions $G$ at some band limit $L\in\mathbb{N}$
on the sphere $S^{2}$, i.e.,  $\widehat{G}_{lm}=0$
for $l\geq L$. From (\ref{27}), the directional correlation of a
band-limited signal $F$ by a band-limited filter $\Psi$, both with
a band limit $L$ on the sphere is thus also band-limited, with the
same band limit: $\widehat{\langle R\Psi\vert F\rangle}_{mn}^{l}=0$
for $l\geq L$.

In practice, the signals $F$ may generally be very precisely approximated as
band-limited, through considerations relative to the physical data acquisition process.
For the typical wavelets described in section \ref{sec:Wavelets},
$\Psi_a$ is also essentially band-limited, to very good approximation,
provided that not too fine scales are considered ($a\nrightarrow 0$). Under these conditions, the wavelet
coefficients of $W_{\Psi}^{F}\left(\rho,a\right)$ can therefore be calculated very precisely,
or even exactly on equi-angular pixelizations, at suitable analysis scales.
This is the scope of the fast directional correlational algorithms discussed in the next two subsections.

We do not consider here the question of the signal reconstruction from its wavelet
coefficients through formula (\ref{5}). The corresponding numerical implementation would require an explicit
discretization of both the scales $a$ and the $SO(3)$ variable $\rho$. First steps in that direction have been undertaken
in \cite{WSbogdanova}.

\subsection{Separation of variables\label{sub:Separation-of-variables}}

The algorithm presented here for the directional correlation is based
on the technique of separation of variables.

The factorized form (\ref{20}) of the spherical harmonics naturally
enables one to compute a direct spherical harmonics transform by separation
of the integrations on the variables $\theta$ and $\varphi$. Conversely,
an inverse transform may be computed as successive summations on the
indices $l$ and $m$, up to the band-limit $L$. Correctly ordering the corresponding operations
provides a calculation of direct and inverse spherical harmonics transforms
in $\mathcal{O}(L^{3})$ operations \cite{SASdriscoll}. This separation
of variables for the spherical harmonics transforms may be performed
on iso-latitude pixelizations on the sphere, i.e.,  pixelizations
for which the sampling in $\theta$ is independent of $\varphi$,
but where the sampling in $\varphi$ may conversely depend on $\theta$.
This is the case for equi-angular pixelizations on the sphere. At
a resolution $L\in\mathbb{N}$, $2L\times2L$ equi-angular pixelizations
are defined by a uniform discretization in $2L$ samples both for
the angles $\theta$ and $\varphi$. Such a pixelization scheme defines
pixels with areas varying drastically with the co-latitude \cite{SASwiaux3}.
In particular, on a $2L\times2L$ equi-angular grid, a sampling result
on the sphere states that the spherical harmonics coefficients of
a band-limited function with band-limit $L$ may be
computed exactly as a finite weighted sum, i.e.,  a quadrature,
of the sampled values of that function \cite{SASdriscoll}. HEALPix
pixelizations (Hierarchical Equal Area iso-Latitude Pixelization)
are also iso-latitude pixelizations, but where the sampling in $\varphi$
explicitly depends on $\theta$. Such a pixelization scheme defines
$12N_{side}^{2}$ pixels of exactly equal areas,
for a resolution parameter $N_{side}=2^{k}$ with $k\in\mathbb{N}$.
The computation of the spherical harmonics coefficients of a band-limited
function is not theoretically exact on HEALPix grids, but can be made extremely precise
by an iteration process  \cite{SASgorski}.
These grids are notably used for the NASA WMAP CMB experiment and the ESA
Planck CMB experiment.

The very same reasoning based on the factorized form (\ref{23}) of
the Wigner $D$-functions enables the calculation the inverse Wigner
$D$-functions transform on $SO(3)$ required by (\ref{26}) in $\mathcal{O}(L^{4})$
operations \cite{SASmaslen1,SASmaslen2}. Considering an iso-latitude
pixelization for the angles $\theta$ and $\varphi$ on the sphere,
the separation of variables for the Wigner $D$-functions transforms
may be performed for any structure of the sampling in the third Euler
angle $\chi$, potentially depending on $\theta$ and $\varphi$.
In particular, at a resolution parameter $L\in\mathbb{N}$, one may
consider a uniform discretization in $2L$ samples for $\chi$. Combined
for example with an equi-angular pixelization at the same resolution
for the angles $\theta$ and $\varphi$ on the sphere, this defines
a $2L\times2L\times2L$ equi-angular sampling in $\rho=(\varphi,\theta,\chi)$
on $SO(3)$.

Consequently, the algorithmic structure based on the separation of
variables on iso-latitude pixelizations on the sphere may be summarized
as follows.\cite{SASkostelec2,SASwiaux2} 
\textbf{(a)} Direct spherical harmonics transforms, $\widehat{\Psi}_{ln}$ and
$\widehat{F}_{lm}$: $\mathcal{O}(L^{3})$. \textbf{(b)} Correlation
$\widehat{\langle R\Psi\vert F\rangle}_{mn}^{l}$ in harmonic
space through (\ref{27}): $\mathcal{O}(L^{3})$. \textbf{(c)}
Inverse Wigner $D$-functions transform $\langle R(\rho)\Psi\vert F\rangle$
on $SO(3)$ through (\ref{26}): $\mathcal{O}(L^{4})$.
The global asymptotic complexity associated with the directional correlation
is thus reduced from $\mathcal{O}(L^{5})$ to $\mathcal{O}(L^{4})$
thanks to the separation of variables. For band-limited signals and filters,
the numerical precision of the algorithm is simply driven by the precision of computation of
the spherical harmonics coefficients. It is therefore very precise on HEALPix grids notably, and
theoretically exact on equi-angular pixelizations.

\subsection{Factorization of rotations\label{sub:Factorization-of-rotations}}

The following algorithm for the directional correlation is based on
the technique of factorization of the three-dimensional rotations.

The three-dimensional rotation operators $R(\rho)$ on functions in
$L^{2}(S^{2},d\Omega)$ on the sphere may be factorized as \cite{SASrisbo,SASwandelt,SASMcewen2}\begin{equation}
R\left(\varphi_{0},\theta_{0},\chi\right)=R\left(\varphi_{0}-\frac{\pi}{2},-\frac{\pi}{2},\theta_{0}\right)R\left(0,\frac{\pi}{2},\chi+\frac{\pi}{2}\right).\label{28}\end{equation}
 The directional correlation relation (\ref{26}) and the expression
(\ref{23}) of the Wigner $D$-functions, matrix elements of the operators
$R(\rho)$, therefore give an alternative expression for the directional
correlation of arbitrary signals $F$ and filters $\Psi$ on the sphere.
We get indeed\begin{equation}
\langle R\left(\rho\right)\Psi\vert F\rangle=\sum_{m,m',n\in\mathbb{Z}}\widehat{\langle R\Psi\vert F\rangle}_{mm'n}e^{i(m\varphi_{0}+m'\theta_{0}+n\chi)},\label{29}\end{equation}
with the Fourier coefficients given by\begin{equation}
\widehat{\langle R\Psi\vert F\rangle}_{mm'n}=e^{i(n-m)\pi/2}\sum_{l\geq C}d_{m'm}^{l}\left(\frac{\pi}{2}\right)d_{m'n}^{l}\left(\frac{\pi}{2}\right)\widehat{\Psi}_{ln}^{*}\widehat{F}_{lm},\label{30}\end{equation}
where $C=\max(\vert m\vert,\vert m'\vert,\vert n\vert)$, and with
the symmetry relation $d_{m'm}^{l}(\theta)=d_{mm'}^{l}(-\theta)$
\cite{SASvarshalovich}.

For a band-limited signal $F$ and a band-limited filter $\Psi$ with
band limit $L\in\mathbb{N}$ on the sphere one has $\vert m\vert,\vert m'\vert,\vert n\vert\leq l<L$.
The factorized form of the imaginary exponentials enables the calculation
of the inverse three-dimensional imaginary exponentials transform
required by (\ref{29}) in $\mathcal{O}(L^{4})$ operations. Just
as for the Wigner $D$-functions transforms, considering an iso-latitude
pixelization for the angles $\theta$ and $\varphi$ on the sphere,
the separation of variables for the three-dimensional imaginary exponentials
may be performed for any structure of the sampling in the third Euler
angle $\chi$. In these terms, the directional correlation algorithm
implemented on iso-latitude pixelizations for the angles $\theta$
and $\varphi$ on the sphere through the factorization of rotations
is structured as follows.
\textbf{(a)} Direct spherical harmonics transforms, $\widehat{\Psi}_{ln}$ and
$\widehat{F}_{lm}$: $\mathcal{O}(L^{3})$.
\textbf{(b)} Correlation $\widehat{\langle R\Psi\vert F\rangle}_{mm'n}$ in harmonic
space through (\ref{30}): $\mathcal{O}(L^{4})$.
\textbf{(c)} Inverse transform $\langle R(\rho)\Psi\vert F\rangle$ through (\ref{29}):
$\mathcal{O}(L^{4})$.
The global asymptotic complexity associated with the directional correlation
is thus also reduced from $\mathcal{O}(L^{5})$ to $\mathcal{O}(L^{4})$
thanks to the factorization of rotations
\footnote{Notice that, while the Euler angles $\varphi_{0}$ and $\chi$ are
in the range $\varphi_{0},\chi\in[0,2\pi[$, the original range for
$\theta_{0}$ is $\theta_{0}\in[0,\pi]$, in order to cover the parameter
space of $SO(3)$. Considering also $\theta_{0}\in[0,2\pi[$ puts
the result on the parameter space of the three-torus $\mathbb{T}^{3}$,
which covers twice the parameter space of $SO(3)$. In that context,
the relation (\ref{29}) is understood as a three-dimensional inverse
Fourier transform, which can be calculated in $\mathcal{O}(L^{3}\log_{2}L)$
operations on a $2L\times2L\times2L$ equi-angular grid on $SO(3)$
by the use of the standard Cooley-Tukey fast Fourier transform algorithm.
This optimization however does not reduce the overall asymptotic complexity
for the directional correlation, still driven by the computation of
(\ref{30}) in $\mathcal{O}(L^{4})$ operations.}.
Again, for band-limited signals and filters,
the numerical precision of the algorithm is simply driven by the precision of computation of
the spherical harmonics coefficients.

\subsection{Optimization with steerable and axisymmetric filters\label{sub:Optimization-with-steerable}}

In terms of our rough estimations of subsection \ref{sub:A-priori-computation},
the separation of variables reduces the computation times on a standard
computer from years to days for the directional correlation of band-limited
signals and filters with band-limit $L\simeq10^{3}$, typically sampled
on megapixels maps. However, as already discussed,
if a large number of simulations have to be analyzed, and at various
scales of the filter, $\mathcal{O}(L^{4})$ calculations remain hardly
affordable even through the use of multiple computers.

Steerable filters are typically considered with a small number of
interpolating functions $M$ (see relation (\ref{12})), that is also
a small azimuthal band-limit $N$ (see relation (\ref{13})) relative
to $L$. The use of such steerable filters further reduces the asymptotic
complexity for the directional correlation. On the one hand, the directional
correlation with a steerable filter $\Psi$ reduces to a $M$-terms
linear combination of standard correlations with the corresponding
basis filters $\Psi_{m}$. For $M \ll L$, the asymptotic complexity
of a directional correlation reduces to that of a standard correlation,
with an \emph{a priori} $\mathcal{O}(L^{4})$ complexity, to which
is simply added the $\mathcal{O}(L^{3})$ linear combination which arises from (\ref{12}).
On the other hand, on iso-latitude pixelizations on the sphere, either
the technique of separation of variables, or the factorization of
three-dimensional rotations can be applied to the standard correlation,
by setting $\chi=0$ in the relations (\ref{26}) or (\ref{29}) respectively.
For a steerable filter with a small azimuthal band limit $N \ll L$,
the Fourier index $n$, with $\vert n\vert<N$, can be excluded from
asymptotic complexity counts. It readily appears that the corresponding
asymptotic complexity for the two algorithms hence reduces to $\mathcal{O}(L^{3})$,
on iso-latitude pixelizations on the sphere.\footnote{Let us remark that the issue of the sampling in $\chi$ is not relevant
for steerable filters. The proper rotations by $\chi\in[0,2\pi[$
are indeed analytically driven, and thus with infinite precision,
by the relation (\ref{12}).}
 At $L\simeq10^{3}$,
our rough estimation of computation times is reduced from years to
tens of seconds. This renders the computation easily affordable, even
when multiple signals and multiple scales are considered.

Details on the algorithmic structure, computation times, memory requirements,
and numerical stability of the corresponding implementations on HEALPix
and equi-angular grids on the sphere may be found in \cite{SASMcewen2}
for the factorization of rotations, and in \cite{SASwiaux2} for the
technique of separation of variables and the optimization with steerable
filters. Notice in that regard that a further optimization of the
algorithm based on the separation of variables and with steerable
filters may be achieved on equi-angular pixelizations on the sphere.
It relies on the fact that Wigner $D$-functions transforms may be
decomposed into linear combinations of spherical harmonics transforms,
which therefore drive the overall asymptotic complexity for the directional
correlation. On $2L\times2L$ equi-angular pixelizations, these spherical
harmonics transforms may be computed in $\mathcal{O}(L^{2}\log^{2}L)$
operations through the Driscoll and Healy algorithm \cite{SASdriscoll,SAShealy1,SAShealy2},
if the associated Legendre polynomials are pre-calculated. As discussed
above, the sampling theorem on equi-angular pixelizations on the sphere
also renders the calculation exact.

The axisymmetry of a filter $A(\theta)$ on the sphere is an extreme case of the steerability, for an azimuthal
band limit $N=1$: $\widehat{A}_{ln}=0$ for $\vert n\vert\geq1$.
In that case, we already emphasized that the proper rotation by $\chi$ has no effect on the
filter and the directional correlation reduces to a standard correlation.
At each scale, the wavelet coefficients of a signal with an axisymmetric
filter therefore live on the sphere $S^{2}$ rather than on $SO(3)$.
The directional correlation relation (\ref{27}) consequently reduces
to the following standard form, giving the spherical harmonics coefficients
$\widehat{\langle R_{0}A\vert F\rangle}_{lm}$ of the correlation
of a signal $F$ with an axisymmetric filter $A$ as the pointwise
product between the filter's Legendre coefficients $\widehat{A}_{l}$,
and the spherical harmonics coefficients of the signal $\widehat{F}_{lm}$:\begin{equation}
\widehat{\langle R_{0}A\vert F\rangle}_{lm}=2\pi\widehat{A}_{l}^{*}\widehat{F}_{lm}.\label{33}\end{equation}
The correlation of a band-limited signal with a band-limited axisymmetric filter
($\widehat{A}_{l}=0$ for $l\geq L$) is therefore readily computed in the harmonic space of $S^{2}$.
On iso-latitude pixelizations on the sphere, the direct
spherical harmonics transform of the signal, and the inverse spherical
harmonics transform of the correlation, can simply be computed by
separation of variables in the spherical harmonics. This provides
an algorithmic structure with $\mathcal{O}(L^{3})$ asymptotic complexity, which again
can be reduced to $\mathcal{O}(L^{2}\log^{2}L)$ on equi-angular pixelizations.

\section{Conclusion\label{sec:Conclusion}}

A new field of complex data processing has emerged in many areas of
science. Scalar and tensor data, often distributed on nontrivial
manifolds, come up at continually increasing resolutions. Powerful
signal analysis techniques need to be developed to process such datasets. 

In this paper, we first reviewed recent formal developments for the
continuous wavelet decomposition of signals on the sphere. Second, we
detailed advances in the definition of the corresponding fast
directional correlation algorithms.

These generic developments can find many applications in various fields
such as computer vision (omnidirectional cameras, ...), biomedical
imaging (functional magnetic resonance imaging, ...), geophysics (signals
on the Earth's surface, ...), or astrophysics and cosmology (signals
on the celestial sphere, ...). In that regard, the important results already
obtained in cosmology through the wavelet analysis of the cosmic microwave
background strongly illustrate the fact that the formalism developed
represents a powerful tool for complex data processing on the sphere \cite{JFAAvielva}.

\section*{Acknowledgments}

The authors wish to thank L. Jacques for the generation of pictures.
Y. W. acknowledges support of the Swiss National Science Foundation
(SNF) under contract No. 200021-107478/1. He is also postdoctoral
researcher of the Belgian National Science Foundation (FNRS). P. V.
was supported by a I3P postdoctoral contract from the Spanish National
Research Council (CSIC), and by the Spanish MEC project ESP2004-07067-C03-01.

{\footnotesize
\centerline{\rule{9pc}{.01in}}
\bigskip
\centerline{Received October 6, 2006}
\medskip
\centerline{Revision received March 14, 2007}
\medskip
\centerline{Signal Processing Institute,  Ecole Polytechnique F\'ed\'erale de Lausanne (EPFL)}
\centerline{CH-1015 Lausanne, Switzerland}
\centerline{e-mail: yves.wiaux@epfl.ch}
\medskip
\centerline{Astrophysics Group, Cavendish Laboratory, University of Cambridge}
\centerline{CB3 0HE Cambridge, United Kingdom}
\centerline{e-mail: mcewen@mrao.cam.ac.uk}
\medskip
\centerline{Instituto de F\'isica de Cantabria (CSIC-UC), E-39005 Santander,  Spain, and}
\centerline{Astrophysics Group, Cavendish Laboratory, University of Cambridge, CB3 0HE Cambridge, United Kingdom}
\centerline{e-mail: vielva@ifca.unican.es}
}

\end{document}